# Contemporary formation of early solar system planetesimals at two distinct radial locations

A. Morbidelli[1], K. Baillié[2], K. Batygin[3], S. Charnoz[4], T. Guillot[1], D.C. Rubie[5], T. Kleine[6,7]

[1]Laboratoire Lagrange, Université Cote d'Azur, CNRS, Observatoire de la Cote d'Azur, 06304 Nice, France - ORCID: 0000-0001-8476-7687 (AM) and 0000-0002-7188-8428 (TG)
[2] IMCCE, Observatoire de Paris, PSL Research University, CNRS, Sorbonne Universités, UPMC Univ Paris 06, Univ. Lille, 75014 Paris, France -ORCID : 0000-0002-2120-6388
[3]Division of Geological and Planetary Sciences California Institute of Technology, Pasadena, CA 91125, USA – ORCID : 0000-0002-7094-7908
[4]Université de Paris, Institut de Physique du Globe de Paris, CNRS,75005 Paris, France – ORCID: 0000-0002-7442-491X
[5]Bayerisches Geoinstitut, Universität Bayreuth, 95440 Bayreuth, Germany – ORCID: 0000-0002-3009-1017
[6]Institut für Planetologie, University of Münster, 48149 Münster, Germany – ORCID: 0000-0003-4657-5961
[7]Max Planck Institute for Solar System Research, Justus-von-Liebig-Weg 3, Göttingen, Germany

**The formation of planetesimals is expected to occur via particle-gas instabilities that concentrate dust into self-gravitating clumps[1-3]. Triggering these instabilities requires the prior pile-up of dust in the protoplanetary disk[4,5]. Until now, this has been successfully modeled exclusively at the disk's snowline[6-9], whereas rocky planetesimals in the inner disk were obtained only by assuming either unrealistically large particle sizes[11,12] or an enhanced global disk metallicity[13]. However, planetesimal formation solely at the snowline is difficult to reconcile with the early and contemporaneous formation of iron meteorite parent bodies with distinct oxidation states[14,15] and isotopic compositions[16], indicating formation at different radial locations in the disk. Here, by modeling the evolution of a disk with ongoing accretion of material from the collapsing molecular cloud[17-19], we show that planetesimal formation may have been triggered within the first 0.5 million years by dust pile-up at both the snowline (at ~5 au) and the silicate sublimation line (at ~1 au), provided turbulent diffusion was low. Particle concentration at ~1 au is due to the early outward radial motion of gas[20] and is assisted by the sublimation and recondensation of silicates[21,22]. Our results indicate that, although the planetesimals at the two locations formed about contemporaneously, those at the snowline accreted a large fraction of their mass (~60%) from materials delivered to the disk in the first few $10^4$ yr, whereas this fraction is only 30% for the planetesimals formed at the silicate line. Thus, provided that the isotopic composition of the delivered material changed with time[23], these two planetesimal populations should have distinct isotopic compositions, consistent with observations[16].**

The goal of this work is to identify the conditions that may lead to the contemporary formation of iron meteorite parent bodies at two distinct radial locations in the disk; one of these locations has to be characterized by a higher temperature than the snowline, so to form ice-free planetesimals. Our model is similar to that in Ref.[9,17-19,24], but comprises a viscosity parameter $α$ that, instead of being held fixed, is reduced from $10^{-2}$ to $5×10^{-4}$ as the accretion rate of mass onto the disk, the disk's local temperature and propensity to undergo gravitational instabilities decrease (see Methods). With this improvement, the early viscous, radially spreading disk evolves over time towards a low-viscosity state, consistent with observations of the dust distribution in protoplanetary disks[25,26].

In our simulations, the Sun starts with half of its current mass (M☉), consistent with a Class-0 protostar, and material is delivered to the Sun-disk system at a rate decaying as *exp(-t/0.1Myr)*. The time-integrated infall of material brings the Sun-disk system to 1M☉ in a few $10^5$ years, a timescale comparable to that in Ref.[17-19] but significantly shorter than in Ref.[9]. The Sun is assumed to accrete the material that falls directly within 0.05 au or is transported by the disk to within this limit. Previous work[9,17-19,24] assumed that the angular momentum of infalling material increases rapidly with time, but modern magneto-hydrodynamical simulations highlight the importance of magnetic breaking in removing angular momentum from the infalling material[27]. Hence, we test different parametrizations of the time-evolution of the effective distance where material falls onto the disk, known as the *centrifugal radius* (Methods). We find that, as long as the inflow of infalling material is vigorous, the radial velocity of the gas is positive (i.e. directed away from the star) beyond the centrifugal radius, whereas, when the inflow wanes, the disk rapidly becomes an accretion disk with a negative radial velocity in its inner part. Because a positive radial velocity of the gas can help in trapping dust particles[20] we look for disks that have a protracted phase of radial expansion in their inner part. Assuming a centrifugal radius decreasing as $R_c=0.35$au$/(M_{sun}(t))^{0.5}$, we obtain a disk that expands radially beyond 0.4 au during the first 0.3 Myr (Fig.1). The time-evolution of the disk temperature is also shown in Fig. 1, whereas the evolution of the surface density and viscosity are depicted in Extended Data Fig. 1 and 2.

In our model, all elements heavier than H and He are injected into the disk together with the gas. They are assumed to be in solid form (dust) when the local temperature of the disk is below their condensation temperature $T_{cond}$. For simplicity we first consider only two broad species: rocks ($T_{cond}$=1,400K) and water ice ($T_{cond}$=170K). Initially, the dust is μm in grain size and is transported outwards during the radial expansion of the disk, while also growing on a timescale proportional to the local dust/gas mass ratio and orbital period[11] (see Methods). We cap the maximal dust grain size to be 10cm beyond the snowline and 5mm within the snowline, in agreement with earlier studies on dust coagulation, bouncing and fragmentation[28]. When dust drifts inwards across the snowline, we assume that the ice sublimates and the remaining 70% of the solid mass is redistributed in 5mm grains[6,7] (Methods). The diffusion of water vapor and its recondensation enhances the solid/gas density ratio at the disk's midplane *beyond* the snowline (Fig. 2a), as found previously[7,8,12].

Inward of the snowline, the solid particles drift towards the Sun until their radial velocity becomes positive because their entrainment in the radially expanding gas dominates over the headwind drag[20]. A small pressure bump also appears, and so particles pile-up just outward of this location (Fig. 2a). However, turbulent diffusion, characterized by the coefficient $D=\nu/$Sc, where $\nu$ is the gas viscosity and Sc is the Schmidt number, smooths the radial distribution and impedes efficient settling towards the midplane of such small particles, even for Sc=10. Thus, a solid/gas volume-density ratio of order unity, as required to trigger the streaming instability[11], is never achieved (Fig. 2a). For this to occur Sc=100 is needed, as assumed in Ref.[9], but such a large value has never been observed in hydrodynamical simulations.

The situation changes if silicate sublimation is also taken into account. We now consider three broad species: refractories ($T_{cond}$=1,400K), silicates ($T_{cond}$=1,000K) and water ice (see Methods) We assume that at T=1,000K half of the rocky mass sublimates and the grains break into mm-size particles of more refractory material. This change in particle size makes the radial flow of solids convergent at the silicate-sublimation front, up to 0.35 Myr (Fig. 2b). Moreover, diffusion and re-condensation of the silicate vapor increase the density of solids beyond the sublimation front. Altogether, this creates a local strong enhancement of the solid/gas ratio in the disk's midplane even for Sc=10. When this ratio becomes larger than[29] 0.5, we convert at each timestep part of the solid density excess into planetesimals[11] (see Methods).

Fig. 3 shows the radial mass distributions of the planetesimal populations produced at the snowline and silicate-sublimation line as a function of time. About 4.5 Earth masses ($M_\oplus$) of silicate-rich planetesimals form in a ring extending from 0.75 to 0.9 au, during a time period from 0.33 to 0.38 Myr, whereas ~32 $M_\oplus$ of ice-rich planetesimals form beyond the snowline, from ~3 to 5.5 au, during 0.1-0.5 Myr. Such a large mass in icy planetesimals can explain the rapid formation of Jupiter's core at the snowline[16] while the concentration of rocky planetesimals in a narrow ring is needed to explain the small masses of Mercury and Mars relative to Earth and Venus[30]. Interestingly, if we further reduce the minimum value of the viscosity parameter α to $10^{-4}$ (instead of $5 \times 10^{-4}$) the total mass of planetesimals produced at the silicate sublimation line exceeds 40 $M_\oplus$. This planetesimal mass, although too large for the Solar System, could readily explain the formation of rocky Super-Earths, which are frequently observed around other stars[31], but are difficult to produce starting from a uniform distribution of planetesimals throughout the disk[32]. Our model predicts that the formation of rocky planets should always be accompanied by the formation of more distant icy planets (Extended Data Fig. 3).

We now compare our results with the constraints from the meteorite record. Iron meteorites are fragments of the metallic cores of some of the oldest planetesimals of the solar system, which formed within 1 million years (Myr) after solar system formation (as defined by the time of formation of its first solids, Ca-Al-rich inclusions or CAIs)[16]. The iron meteorites can be subdivided into two isotopically distinct groups, which are termed the carbonaceous (CC) and non-carbonaceous (NC)[16] groups. Of note, the parent bodies of the CC irons tend to have smaller relative core sizes and are characterized by lower Fe/Ni ratios than those of the NC irons (see Methods), suggesting that the former formed in more oxidizing environments than the latter. As such, our working hypothesis is to identify the planetesimals formed at the snowline as the parent bodies of CC iron meteorites, consistent with their formation in a more oxidizing environment, and those formed at the silicate-sublimation line as the parent bodies of NC irons, consistent with the observation that they accreted at higher temperature and were water-ice free. A larger water-ice fraction in CC iron parent bodies also leads to a more protracted timescale of core formation, due to the lower concentration of heat-producing $^{26}$Al[15]. This is consistent with the observed later core formation time of CC compared to NC iron parent bodies at ~3 Myr and ~1 Myr, respectively[15,16].

Another important difference between planetesimals formed at the silicate-sublimation line and at the snowline is that our model predicts the former to have silicate (olivine + pyroxene)-to-refractory-element ratios 10-35% higher than the proto-solar value. This property results from re-condensation of the gas that sublimated off refractory grains at high temperature[21,22,33]. The Ni/Ir inferred for most bulk NC cores is indeed larger than solar[14] (Ni condenses together with silicates, whereas Ir is refractory), but this property is not unique to NC irons[34,35]. So, these data do not provide clear evidence for the formation of NC parent bodies at the silicate-sublimation line. Instead, the enhanced silicate-to-refractory element ratio predicted by our model is consistent with the supra-solar Si/Al ratios of NC chondrites, which are not observed in any CC chondrites. Chondrites are later-formed planetesimals, and so modeling their formation goes beyond the scope of this study (see Supplementary Section S5 for a discussion). Nevertheless, early-formed planetesimals may well be their precursors via the subsequent generation of chondrules as collisional debris[36]. Consequently, the chemical composition of

NC chondrites may still reflect that of the first planetesimals formed at the silicate-sublimation line, modeled in this work.

The most important constraint is that of the aforementioned isotopic dichotomy between NC and CC irons[16]. To test the ability of our model to satisfy this constraint, we distinguish between material accreted to the disk before and after the first 20 Kyr (denoted "early" and "late material" hereafter; Fig. 3). The choice of this time is justified in the Supplementary Section S1.5 and the relationship between the condensates of the early material and CAIs is discussed in Supplementary Section S4. We find that planetesimals formed at the snowline incorporate a larger fraction of early material than planetesimals at ~1au (Fig. 3). This is because early material is efficiently transported to the outer disk during the radial expansion phase, while it is substituted by late-infalling material in the inner disk. By the time the inner planetesimals form, the drift of early material back into the inner disk again raises the early-to-late material ratio at ~1au, but this ratio nevertheless remains below that of the outer disk (Fig. 4). Assuming that the early and late materials are isotopically distinct[23], the two populations of planetesimals produced in our model at distinct radial locations have distinct isotopic compositions, as observed for NC and CC irons[16]. Moreover, the mixing ratios between early and late materials in our model are in good agreement with those derived from the isotopic offset between the NC and CC reservoirs (see Methods). Finally, we note that, although the presence of a barrier against dust drift is not needed to explain the isotopic dichotomy between the two populations of early-formed planetesimals modeled in this study, it is nevertheless needed before the disk is completely homogenized, because otherwise the NC-CC dichotomy could not be preserved for the later-formed parent bodies of chondrites. The formation of Jupiter from the population of ice-rich planetesimals (not included in our model) would be the most obvious cause of the appearance of such barrier[16].

Our model highlights the fundamental processes and properties needed to act in concert to account for the meteoritic evidence for the contemporaneous formation of two isotopically-different planetesimal populations at distinct radial locations: (i) a small centrifugal radius for the material falling onto the disk, which is necessary to sustain a protracted radial expansion of the gas and delay the inward drift of dust particles into the Sun; (ii) sublimation and recondensation of water and silicates at the respective phase-transition lines, together with stepwise changes in the maximal sizes of solid particles at each line, to enhance the local solid/gas ratio; (iii) a reduced turbulent diffusion, allowing for sufficient particle pile-up and sedimentation towards the mid-plane, together with a quite large disk temperature so that the silicate sublimation line is initially near 1 au; (iv) a rapid change in isotopic composition of the material accreted onto the disk over time, to account for the radial isotopic gradient in the disk that results in the NC-CC dichotomy when planetesimals form in non-contiguous regions. Importantly, within the context of the proposed model, any derogation from (i)-(iv) would lead to results inconsistent with the meteorite record (Supplementary Sections S1 and S2).

## Methods

### Code description

Structure. Our code uses a one-dimensional grid, similar to Ref.[12,24,37], describing the radial distribution of gas and dust and their properties. The grid samples a user-defined radial range (from $r_{min}$=0.05au to $r_{max}$=100 au for the simulation presented in the main text) in logarithmic bins. We used 100 bins for the presented simulation, although different numbers of bins have been used in convergence tests.

Accretion of mass on the disk. For numerical reasons, the disk is initialized with an arbitrarily small surface density and a temperature $T_{irr}$=115K $(r/au)^{-3/7}$, corresponding to a passively irradiated disk[38]. The gas is supplied at a rate

$$\dot{M}(t) = \frac{M_\odot - M_{sun}(0)}{\tau} \exp\left[-\frac{t}{\tau}\right] \qquad (1)$$

where $M_{sun}(0)$ is its initial mass in the simulation (here ½ $M_\odot$). Previous work[17-19] assumed that mass infall rate (1) is constant, but truncated the infall abruptly when the mass of the star-disk system reached 1 $M_\odot$. We think it is more realistic to expect that the accretion rate decays over time, with a low-rate tail and no artificial truncation. We adopt $\tau = 10^5$y, which is of the order of the duration of the infall in Ref.[13,17,18] (170 Kyr). The gas falling in a radial bin is[17-19,24]:

$$\dot{M}(r) = \left[\left(1 - \sqrt{r_-/R_c(t)}\right)^{1/2} - \left(1 - \sqrt{r_+/R_c(t)}\right)^{1/2}\right] \dot{M}(t) \qquad (2)$$

where $r_+$ and $r_-$ are the upper and lower boundary of each bin. $R_c(t)$ is called *centrifugal radius* or *injection radius* and its parametrization is given in input. In previous studies[17-19] $R_c$ was assumed to grow as $R_c(t)$=10au $(M_{sun}(t)/M_\odot)^3$ but in this work we test different parametrizations. The nominal simulation presented in the main text is obtained with $R_c(t)$=0.35au/$(M_{sun}(t))^{0.5}$. The effect of this $R_c$ prescription is discussed in Supplementary Section S1.1.

Computation of the disk temperature. The midplane temperature $T$ in each ring of the disk, is computed taking into account several contributions. The first is the energy released by the infalling material shocking at the surface of the considered disk ring:

$$Q_{infall} = \frac{1}{2}\frac{GM_{sun}(t)\dot{M}(r)}{r} \qquad (3)$$

where $G$ is the gravitational constant. We take the conservative assumption that only ½ of the final potential energy of the infalling gas is injected in the disk (hence the factor ½ in (3)), the rest being lost during the infalling phase. The second contribution is the energy released by viscous heating[39]:

$$Q_{visc} = 2\pi r \delta r \frac{9}{4}\Sigma_g \nu \Omega^2 \qquad (4)$$

where $\Sigma_g$ is the surface density of the gas in the ring of width $\delta r = (r_+ - r_-)$, $\nu$ is the viscosity, and $\Omega$ is the Keplerian frequency. The third contribution is the ring's cooling due to black body irradiation at its surfaces:

$$Q_- = 2 \times 2\pi r \delta r \sigma_B T_s^4 \qquad (5)$$

where $\sigma_B$ the Stephan-Boltzman constant and $T_s$ is the temperature at the surface of the disk, related to the midplane temperature $T$ by the relationship[39]:

$$T_s^4 = \frac{4}{3}\frac{2T^4}{\kappa \Sigma_g} \qquad (6)$$

which is valid where the disk is optically thick. The opacity $\kappa$ is a function of temperature[40]. The last contribution is that of energy exchange between adjacent disk's rings. A ring gains or loses energy at a rate $\delta F=F_+-F_-$, where $F_+$ (resp. $F_-$) is the flux of energy across the boundary with the external (resp. internal) adjacent ring[41]:

$$F = (2\pi)^{3/2} \frac{16\lambda\sigma_B}{\kappa\rho_g} \frac{dT}{dr} T^3 r H \qquad (7)$$

where $\rho_g=\Sigma_g/[(2\pi)^{1/2}H]$ is the volume density of the gas, $H=(\mathcal{R}/\mu \, Tr)^{1/2}$ is the pressure scale-height of the disk ($\mathcal{R}$ being the gas constant and $\mu$ the gas mean molecular weight), $\lambda$ is the flux-limiter[42] and all quantities are taken at the boundary between adjacent rings. For $\mu$ we assume 2.3g/mol and we take the approximation to keep this number constant across condensation lines. The quantity $(Q_{infall}+Q_{visc}-Q_-+\delta F)\delta t$ describes the change of internal energy of a ring over an integration timestep $\delta t$ and the change of temperature $T$ is obtained by dividing this quantity by the heat capacity of the ring $c_v= \mathcal{R}/[(\gamma-1)\mu]$, and $\gamma=1.4$ is the adiabatic index. If the temperature falls below that of a passively irradiated disk $T_{irr}$, we reset $T=T_{irr}$. The temperature is further modified during the advection step, described below.

Viscosity prescription. The viscosity $\nu$ is as usual defined as $\nu=\alpha H^2\Omega$. The viscosity parameter $\alpha$ had been set constant and equal to $10^{-2}$ in previous works[17-19]. In this work we change $\alpha$ over time and radial location. We set:

$$\alpha = \alpha_{min} + (\alpha_{max} - \alpha_{min})\frac{\dot{M}(t)}{\dot{M}(0)} \qquad (8)$$

the rationale being that the infall of material onto the disk generates Reynolds stresses that act as a viscosity[27], which become weaker as the infall wanes. It is also known that at high temperature, typically above the silicate sublimation value, the disk becomes prone to ionization and to the magneto-rotational instability, which raises the turbulent viscosity significantly. Thus, for the rings with temperature $T>1,500$K we set $\alpha=\alpha_{max}$ and for rings with $1,000$K $< T <$ $1,500$K we set $\alpha$ to a $T$-dependent value intermediate between (8) and $\alpha_{max}$ computed as:

$$f = \sin\left[\frac{T-1,000}{1,000}\pi\right], \quad \alpha(T) = \exp[(1-f)\log\alpha + f\log\alpha_{max}]$$

Similarly, it is known that when the disk is gravitationally unstable or close to instability, the disk develops clumps and waves that also generate an effective viscosity[43]. Thus, for the rings where Toomre's Q parameter[43] is less than unity (a criterion for gravitational instability) we set $\alpha=3\times10^{-2}$ and for rings with $1<Q<Q_{lim}$ we set $\alpha$ to a Q-dependent value intermediate between (8) and $3\times10^{-2}$, given by:

$$f = \sin\left[\frac{Q-1}{2(Q_{lim}-1)}\pi\right], \quad \alpha(T) = (1-f)3\times 10^{-2} + f\alpha$$

For the nominal simulation presented in the main text we set $\alpha_{max}=10^{-2}$, $\alpha_{min}=5\times10^{-4}$, $Q_{lim}=10$. We discuss in the Supplementary Section S1.2 how the results change when these parameters are varied. We find that radial energy exchange (7), not included in previous codes[17-19], is essential to stabilize the disk when $\alpha$ is allowed to change over time at different radii as in our model.

Computation of gas evolution. We now discuss how the surface density of gas in the disk evolves. Besides receiving mass at a rate (2) a ring can exchange material with neighboring rings. The radial velocity of the gas at the boundary between two rings is due to the mutual viscous torques that they arise on each other due to the differential rotation and results:

$$v_r^g = -\frac{3}{\Sigma_g\sqrt{r}}\frac{d}{dr}\left(\Sigma_g\nu\sqrt{r}\right) \qquad (9)$$

where all quantities are evaluated at the boundary. This speed is then modified to account for the back-reaction of dust onto gas, as will be discussed below. A ring gains or loses mass at a rate $\delta F_M = F_{M-} - F_{M+}$, where $F_{M+}$ (resp. $F_{M-}$) is the flux of mass across the boundary with the external (resp. internal) adjacent ring:

$$F_M = 2\pi r v_r^g \Sigma_g \qquad (10)$$

where $\Sigma_g$ is here the surface density in the ring that is supplying mass to the other ring and $r$ is the radial distance of the boundary between the considered rings, where (9) has been computed. From (10) the change in surface density of a ring over an integration timestep is readily computed. For the stability of the code we impose that $\delta t \, v_r^g < 0.2 \delta r$. As anticipated, together with the advection of mass, we also compute an advection of thermal energy associated to the flux of gases in/out of rings with different temperatures. This further modifies the ring's temperature, on top of the prescription described above.

In doing these calculations, the boundary conditions play an important role, in particular the inner one (the outer boundary being very far from the region of interest). We use open boundary conditions, assuming that the gas flowing inward of $r_{min}$ is immediately accreted by the central star along magnetic channels (we increase the stellar mass accordingly) and that the gas flowing beyond $r_{max}$ is immediately photoevaporated or stripped away by passing stars.

Dust species and particle growth. When the gas is supplied to the disk following eq. (2) we also assume that 1% of its mass is supplied in condensable materials. We consider three type of materials: ice, with a condensation temperature of T=170K, silicates, with a condensation temperature of T=1,000K and refractories, with a condensation temperature of T=1,400K. This choice, lower than those canonically assumed for the condensation of silicates and refractory elements, is discussed in Supplementary Section S3. For simplicity we neglect that the condensation temperature depends on the partial pressure of the considered material. Then, we introduce three surface density functions $\Sigma_{ice}$, $\Sigma_{sil}$ and $\Sigma_{ref}$ for these three materials, respectively. At injection, we assume that 30% of the condensable material is in ice, consistent with comet composition, 35% is in silicates and 35% in more refractory materials. When the temperature is larger than the corresponding condensation temperature, the material is considered to be in vapor form, while for smaller temperature it is assumed to have condensed into dust.

The dust has initially a size (diameter) of 1μm, but then grows with a timescale

$$\tau_{gr} = \frac{1}{Z\Omega} \qquad (11)$$

where Z is the local solid/gas surface density ratio (see Ref.[11] for a derivation). For simplicity we consider only one dust size in each ring, instead of a size distribution. The reason is that in dust growth models, most of the dust's mass is concentrated in particles near the maximal dust-size[44]. Because of this simplification, when new dust is created in a bin that already hosts partially-grown dust (for instance due to the injection of fresh material from the molecular cloud) we take the total-mass-weighted mean size between the pre-existing size and the new injected one[45]. The maximum dust-size is set by the drifting, bouncing and fragmentation barriers. Based on previous works on the effect of these barriers[8,44] we limit the maximal size of dust in the ice-regime (T<170K) to 10cm, that in the silicate regime (1,000K>T>170K) to 5mm and that in the refractory regime (1,400K>T>1,000K) to 1mm. Silicate and refractory particles are thus much smaller than in Ref.[11,12], which is consistent with a reduced fragmentation energy[8] than that considered in those works and the existence of a bouncing barrier[28]. These sizes are also consistent at the order of magnitude with those of silicate and refractory particles observed in meteorites, such as chondrules, chondrule clusters and CAIs. The large size-contrast between icy and silicate particles is due to the fact that warm ice (near the snowline, where planetesimal formation will take place) is more sticky than silicates, so aggregates are expected to grow bigger[8]. The size contrast between silicate and refractory particles could be justified by fragmentation during silicate sublimation. We discuss in Supplementary Section S1.6 how the results change with different size contrasts. Once the dust size is set, the dust's Stokes number $S_t$ is computed from the local density of gas[11]. Thus, a particle drifting towards the Sun with constant size has its Stokes number progressively reduced because the disk's gas density increases.

Evolution of the dust surface densities $\Sigma_{ice}$, $\Sigma_{sil}$ and $\Sigma_{ref}$. When the corresponding material is in vapor form, we assume that its radial velocity is equal to that of the disk's gas (9). When it is in dust form, its radial velocity $v_r^d$ is computed as described in the appendix of Ref.[11], which includes a modification of the gas radial velocity (9) due to the back-reaction of dust on gas. Differently from Ref.[11], this modification of the gas velocity affects the evolution of the gas. Thus, our model is able in principle to capture the so-called *self-induced trap* phenomenon[46]. For the record, we never observe this phenomenon in our nominal simulations, because the radial velocity of the gas is positive during the planetesimal-formation stage, but we do observe it in a classic, viscous accretion disk model if the particle size is 10cm.

In addition to the advection process, analog to that of the gas described above (eq. 10 with $\Sigma_d$ and $v_r^d$ instead of $\Sigma_g$ and $v_r^g$, where $\Sigma_d$ stands generically for $\Sigma_{ice}$, $\Sigma_{sil}$ or $\Sigma_{ref}$), the evolution of $\Sigma_d$ is also affected by a diffusion equation[11] which, from the ring perspective followed in this code description, generates a supplementary mass gain/loss $\delta F_D = F_{D-} - F_{D+}$, where $F_{D+}$ (resp. $F_{D-}$) is the flux of mass due to diffusion across the boundary with the external (resp. internal) adjacent ring:

$$F_D = 2\pi r D \Sigma_g \frac{d}{dr}\left(\frac{\Sigma_d}{\Sigma_g}\right) \qquad (12)$$

Here $D$ is the diffusion coefficient and $r$, $\Sigma_g$ and the gradient of $\Sigma_d/\Sigma_g$ are evaluated at the boundary between the adjacent rings. We set $D=\nu/Sc$, Sc being the Schmidt number. For a passive tracer of the gas, Sc can be[47] as large as 10, which is our nominal choice (see Supplementary Section S1.2 for a discussion of the effects of this number). We assume the same Sc for vapor and solid particles because[26] the Stokes number of our particles is always smaller than 0.1 for $r<8$ au and $t<0.5$ Myr.

The use of a unique density function $\Sigma_d$ to describe both the vapor and solid phases of the same material, which just differ in advection radial velocity, is a simplified but effective way to treat the sublimation/recondensation process. If one treats vapor and dust separately, each of their respective density functions has a discontinuity, dropping to zero at the condensation/sublimation line. If, instead of assuming instantaneous sublimation/condensation as we do here for simplicity, one considers a non-zero condensation or sublimation timescale dependent on partial pressures[8,10,12], the discontinuity becomes a gradient, but such a gradient is nevertheless very steep. Consequently, in both cases eq. (12) would give a strong mass flux at the boundary between the vapor-dominated and the dust-dominated regimes. But, in reality, most of the diffusion of vapor through the condensation line is counterbalanced by the diffusion of dust in the opposite direction. If, instead, one uses a single density function for both vapor and dust, as we do here, eq. (12) automatically describes the net mass flux across the sublimation/condensation boundary, which is the one that really matters. Our procedure is mathematically exact if one makes the simplifying assumption of instantaneous and complete sublimation/condensation at the critical temperature. We provide in Supplementary Section S1.7 a test on the expected differences in the results using the two approaches.

Planetesimal formation. Associated to the surface densities of gas ($\sum_g$) and dust ($\sum_d$) there are the volume densities on the midplane $\rho_g=\sum_g/[(2\pi)^{1/2}H]$ and $\rho_d=\sum_g/[(2\pi)^{1/2}H_d]$, with $H_d=H_g[\alpha/Sc/(\alpha/Sc+St)]^{1/2}$. In the calculation of $\rho_d$ we sum up the contributions of all the three species - ice, silicate and refractory. Whenever in a ring $\rho_d/\rho_g>0.5$ we assume that planetesimal formation can take place via the streaming instability in that ring[29]. A fraction 0.01% of the solids is converted into planetesimals per ring's orbital period[11]. A more elaborated prescription[13] is also tested in Supplementary Section S1.3. The corresponding mass is subtracted from the dust density functions in proportion to the relative abundances of the three species of dust. If a species is in vapor form, of course its density remains untouched. When the production of planetesimals reduces the dust/gas mass ratio below 0.5, planetesimal formation is stopped. This regulates planetesimal production and dust accumulation, keeping the dust/gas ratio typically below unity (Fig. 2). Without planetesimal formation the ratio would increase further. Thus, the very phenomenon of planetesimal formation is not very sensitive on the adopted dust/gas threshold ratio, although the resulting total mass of planetesimals does increase/decrease if the adopted threshold is decreased/increased.

Isotopic composition. In order to study the isotopic composition of planetesimals we compute the evolution of populations of dust tracers in the disk. We follow the idea proposed in Ref.[23] according to which the isotopic dichotomy between NC and CC planetesimals is due to the injection in the disk of isotopically distinct materials at different times. Thus, we define a switch time $t_{dich}$ and we split $\sum_{ref}$ into two functions, $\sum_{ref}^{(1)}$ and $\sum_{ref}^{(2)}$ describing the surface density of refractory material injected at $t<t_{dich}$ or at $t>t_{dich}$, respectively. The two distributions are referred to as "early" and "late infalling material" in the main text. Both $\sum_{ref}^{(1)}$ and $\sum_{ref}^{(2)}$ undergo the diffusion process described by equation (12), which induces their mutual mixing, illustrated in Fig. 4. When a planetesimal forms at time $t$, its composition in terms of early vs. late material is given by the $\sum_{ref}^{(1)}/\sum_{ref}^{(2)}$ ratio at that time and location in the disk. Averaging the compositions of planetesimals formed in the same radial bin at different times (mass-weighted average) produces Fig. 3.

**Determining the sizes of cores of NC and CC iron meteorites parent bodies**

The core sizes of iron meteorite parent bodies can be estimated using the abundances of highly siderophile elements (HSEs) inferred for the bulk cores. Owing to their strong siderophile character, the HSEs quantitatively partitioned into the core. Thus, the core mass fraction can be calculated by dividing the HSE concentrations of the bulk body (assumed to be chondritic) by the HSE concentrations of the bulk core. To this end, the HSE concentrations of the bulk cores are inferred by modelling fractional crystallization. For the refractory HSEs Re, Os, Ir, Ru, and Pt the resulting relative ratios are typically broadly chondritic, such that for each element similar core mass fractions are calculated. The HSE concentration data used for calculating core mass fractions are summarized in Extended Data Tables 1 and 2 summarizes the mean core mass fractions for each iron meteorite parent body. For the CC irons we assumed that, prior to core formation, the bulk body had a CI-chondritic composition. This is the most appropriate composition, given that these bodies formed at the snow line and, therefore, incorporated water ice. For the NC irons we used an average ordinary (OC) or enstatite chondrite (EC) composition. However, using the same starting compositions for both NC and CC irons does not change the resulting core mass fractions by much, except that it would lead to overlapping values for the IIC and IVA irons.

The calculations reveal overall larger core mass fractions in NC compared to CC iron meteorite parent bodies, where NC cores typically were ~20% of the mass of the parent body, and CC cores were <15% (Extended Data Table 2). The smaller core sizes of the CC parent bodies are consistent with a larger water ice fractions in these bodies, which results in oxidation and, hence, a smaller fraction of Fe partitioned into the core. This is also consistent with the systematically lower Fe/Ni ratios inferred for CC compared to NC cores[15]. These observations are difficult to reconcile with a model[37] where the parent bodies of both NC and CC iron meteorites would have formed at the snowline at different times. By contrast, these observations are fully consistent with our model in which NC and CC bodies formed in rocky and icy environments, respectively.

The existence of water ice in CC iron meteorite parent bodies changes drastically the thermal evolution models usually used to infer the accretion time of planetesimals from their measured differentiation times. Until now, most thermal modelling studies assumed the same composition for NC and CC iron meteorite parent bodies, resulting in a monotonic relationship between accretion time and differentiation time, which in turn implied that CC iron meteorite parent bodies accreted later than their NC counterpart[16]. However, a more recent study showed that water ice delays the onset of melting and core formation[15] and that, therefore, the later core formation time of CC iron parent bodies does not imply later accretion. In fact, once the effect of different water ice fractions is taken into account, the inferred accretion times of CC and NC iron meteorite parent bodies are indistinguishable, where both groups of bodies are constrained to have formed within the first 1 Myr of the solar system[15], in line with our predictions.

**Determining the mixing ratios of distinct materials using the isotopic properties of CAIs, CC and NC meteorites**

For all elements which display the NC-CC isotopic dichotomy, the CC reservoir is always between the isotopic compositions of CAIs and NC meteorites[48]. This observation has led to the proposal that the NC-CC dichotomy reflects different mixing proportions of two isotopically distinct disk reservoirs, which were characterized by similar, broadly chondritic bulk chemical compositions[23,48]. One of these reservoirs is characterized by a CAI-like isotopic composition (termed *IC* for Inclusion-like Chondritic reservoir[48]) and corresponds to the early-infalling material. Note that although this material has a CAI-like *isotopic* composition, its *chemical* composition is distinct from CAIs and instead is assumed to be chondritic. This assumption stems from the observation that the NC-CC dichotomy exists for refractory (e.g., Mo, Ti) and non-refractory elements (e.g., Cr, Ni) and that for all these elements the CC reservoir is isotopically always intermediate between NC and CAIs. The other reservoir is characterized by an NC-like isotopic composition, but its exact isotopic composition, termed $NC_i$, is not known. Within this framework the isotopic composition of the CC and NC reservoirs can be expressed as simple binary mixtures between *IC* and $NC_i$ material as follows:

$$CC = x \cdot IC + (1-x) \cdot NC_i$$
$$NC = y \cdot IC + (1-y) \cdot NC_i$$

where *x* and *y* denote the fractions of early material in the CC and NC reservoir, respectively. These two parameters cannot be calculated independently, because the isotopic composition of the late infall, $NC_i$, is not known. However, the two equations can be combined by eliminating $NC_i$, so that *x* can be calculated as a function of *y* as follows:

$$x = \frac{y(CC - IC) + NC - CC}{NC - IC}$$

The NC-CC dichotomy is best defined for Ti, Cr, and Mo isotopes, which therefore are most suitable to calculate the dependence of $x$ and $y$. As the $\varepsilon^{50}$Ti and $\varepsilon^{54}$Cr isotope anomalies among NC meteorites are correlated, and because this correlation points towards the composition of the CC reservoir, using $\varepsilon^{50}$Ti or $\varepsilon^{54}$Cr returns the same results. For Ti we used the following values: $\varepsilon^{50}$Ti$_{IC}$ = +9, which is the average Ti isotope anomaly of CAIs[49]; $\varepsilon^{50}$Ti$_{NC}$ = –1, which is the average $\varepsilon^{50}$Ti of NC meteorites, or its most extreme negative value $\varepsilon^{50}$Ti$_{NC}$ = –2; the Ti isotopic composition of CI chondrites, $\varepsilon^{50}$Ti$_{CC}$ = +2, which best represent the composition of the outer disk[48]. Using different values for $\varepsilon^{50}$Ti$_{NC}$ or $\varepsilon^{50}$Ti$_{CC}$ within the compositional range of the NC and CC reservoirs does not change the result significantly. For Mo we use the characteristic Mo isotopic difference between the CC and NC reservoirs, which can be expressed as $\Delta^{95}$Mo (see Ref.[50]), with the following values: $\Delta^{95}$Mo$_{IC}$ = +125 (Ref.[49]); $\Delta^{95}$Mo$_{CC}$ = +26 (Ref.[50]); and $\Delta^{95}$Mo$_{NC}$ = -9 (Ref.[50]).

The relation between the fraction of early material in the CC and NC reservoirs $x$ and $y$, respectively, calculated using the Ti and Mo isotope anomalies are shown in Extended Data Fig. 4 together with the proportions predicted by our model for $t_{dich}$ = 20 Kyr (see above). This comparison shows that our model can reproduced the early-to-late material ratios in the NC and CC reservoirs quite well.

**Data availability:** lagrange.oca.eu/images/LAGRANGE/pages_perso/morby/forNature.tar.gz provides the compiled code, the input file and the ascii output files of our reference simulation including silicate condensation/sublimation: one file per output timestep ($10^4$yr) for a total of 100 files. A readme file describes the content of each file

**Code availability:** The source of the code can be provided upon request to the corresponding author

**Acknowledgments:** A.M. and S.C. acknowledge support from program ANR-20-CE49-0006 (ANR DISKBUILD). The work presented here has been performed in preparation of the proposal *HolyEarth* by A.M. and T.K., which has been funded by the ERC (grant N. 101019380). The authors thank the three reviewers, including R. Deienno, for their constructive and detailed comments.

**Author contributions**: A.M. conceived the project, wrote the code, ran the simulations and led the writing of the manuscript. Ke.B., wrote an earlier version of the code. Ke.B., S.C. and T.G. contributed with their experience on disk evolution. Ko.B. stressed the importance of the radial expansion of the disk. D.R. and T.K. provided their experience on the chemical and isotopic composition of meteorites which allowed testing the model against measured constraints. All authors contributed to writing the manuscript and discussing the significance of the results.

**Competing interests**. The authors declare no competing interests

**Correspondence and requests for materials** should be addressed to A.M.

# References

[1] Youdin, A.N., Goodman, J. 2005. Streaming Instabilities in Protoplanetary Disks. The Astrophysical Journal 620, 459–469. Doi :10.1086/426895

[2] Cuzzi, J.N., Hogan, R.C., Shariff, K. 2008. Toward Planetesimals : Dense Chondrule Clumps in the Protoplanetary Nebula. The Astrophysical Journal 687, 1432–1447. Doi :10.1086/591239

[3] Simon, J.B., Armitage, P.J., Li, R., Youdin, A.N. 2016. The Mass and Size Distribution of Planetesimals Formed by the Streaming Instability. I. The role of Self-gravity. The Astrophysical Journal 822. Doi :10.3847/0004-637X/822/1/55

[4] Drążkowska, J., Dullemond, C.P. 2014. Can dust coagulation trigger streaming instability ?. Astronomy and Astrophysics 572. Doi :10.1051/0004-6361/201424809

[5] Yang, C.-C., Johansen, A., Carrera, D. 2017. Concentrating small particles in protoplanetary disks through the streaming instability. Astronomy and Astrophysics 606. Doi :10.1051/0004-6361/201630106

[6] Ida, S., Guillot, T. 2016. Formation of dust-rich planetesimals from sublimated pebbles inside of the snow line. Astronomy and Astrophysics 596. Doi :10.1051/0004-6361/201629680

[7] Schoonenberg, D., Ormel, C.W. 2017. Planetesimal formation near the snowline : in or out ?. Astronomy and Astrophysics 602. Doi :10.1051/0004-6361/201630013

[8] Drążkowska, J., Alibert, Y. 2017. Planetesimal formation starts at the snow line. Astronomy and Astrophysics 608. Doi :10.1051/0004-6361/201731491

[9] Drążkowska, J., Dullemond, C.P. 2018. Planetesimal formation during protoplanetary disk buildup. Astronomy and Astrophysics 614. Doi :10.1051/0004-6361/201732221

[10] Hyodo, R., Ida, S., Charnoz, S. 2019. Formation of rocky and icy planetesimals inside and outside the snow line: effects of diffusion, sublimation, and back-reaction. Astronomy and Astrophysics 629. doi:10.1051/0004-6361/201935935

[11] Drążkowska, J., Alibert, Y., Moore, B. 2016. Close-in planetesimal formation by pile-up of drifting pebbles. Astronomy and Astrophysics 594. Doi :10.1051/0004-6361/201628983

[12] Charnoz, S. and 6 colleagues 2019. Planetesimal formation in an evolving protoplanetary disk with a dead zone. Astronomy and Astrophysics 627. Doi :10.1051/0004-6361/201833216

[13] Schoonenberg, D., Ormel, C.W., Krijt, S. 2018. A Lagrangian model for dust evolution in protoplanetary disks: formation of wet and dry planetesimals at different stellar masses. Astronomy and Astrophysics 620. doi:10.1051/0004-6361/201834047

[14] Chabot, N. L. Composition of metallic cores in the early solar system. 49th Lunar and Planetary Science Conference,  LPI Contrib. No. 2083, (2018).

[15] Spitzer, F., Burkhardt, C,. Nimmo, F., Kleine, T. (2021):       Nucleosynthetic Pt isotope anomalies and the Hf-W chronology of core formation in inner and outer solar system planetesimals. Earth and Planetary Science Letters, in press.

[16] Kruijer, T.S., Burkhardt, C., Budde, G., Kleine, T. 2017. Age of Jupiter inferred from the distinct genetics and formation times of meteorites. Proceedings of the National Academy of Science 114, 6712–6716. Doi :10.1073/pnas.1704461114

[17] Hueso, R., Guillot, T. 2005. Evolution of protoplanetary disks : constraints from DM Tauri and GM Aurigae. Astronomy and Astrophysics 442, 703–725. Doi :10.1051/0004-6361 :20041905

[18] Pignatale, F.C., Charnoz, S., Chaussidon, M., Jacquet, E. 2018. Making the Planetary Material Diversity during the Early Assembling of the Solar System. The Astrophysical Journal 867. Doi :10.3847/2041-8213/aaeb22

[19] Baillié, K., Marques, J., Piau, L. 2019. Building protoplanetary disks from the molecular cloud : redefining the disk timeline. Astronomy and Astrophysics 624. Doi :10.1051/0004-6361/201730677

[20] Batygin, K., Morbidelli, A. 2020. Formation of Giant Planet Satellites. The Astrophysical Journal 894. Doi :10.3847/1538-4357/ab8937

[21] Aguichine, A., Mousis, O., Devouard, B., Ronnet, T. 2020. Rocklines as Cradles for Refractory Solids in the Protosolar Nebula. The Astrophysical Journal 901. Doi :10.3847/1538-4357/abaf47

[22] Miyazaki, Y. and Korenaga, J. 2021. Dynamic evolution of major element chemistry in protoplanetary disks and its implications for Earth-enstatite chondrite connection. Icarus 361. doi:10.1016/j.icarus.2021.114368

[23] Nanne, J.A.M., Nimmo, F., Cuzzi, J.N., Kleine, T. 2019. Origin of the non-carbonaceous-carbonaceous meteorite dichotomy. Earth and Planetary Science Letters 511, 44–54. Doi :10.1016/j.epsl.2019.01.027

[24] Appelgren, J., Lambrechts, M., Johansen, A. 2020. Dust clearing by radial drift in evolving protoplanetary discs. Astronomy and Astrophysics 638. Doi :10.1051/0004-6361/202037650

[25] Pinte, C. and 6 colleagues 2016. Dust and Gas in the Disk of HL Tauri : Surface Density, Dust Settling, and Dust-to-gas Ratio. The Astrophysical Journal 816. Doi :10.3847/0004-637X/816/1/25

[26] Dullemond, C.P. and 14 colleagues 2018. The Disk Substructures at High Angular Resolution Project (DSHARP). VI. Dust Trapping in Thin-ringed Protoplanetary Disks. The Astrophysical Journal 869. Doi :10.3847/2041-8213/aaf742

[27] Lee, Y.N., Charnoz, S., Hennebelle, P. 2021. Protoplanetary disk formation from the collapse of a prestellar core. Astronomy and Astrophysics 648

[28] Güttler, C., Blum, J., Zsom, A., Ormel, C.W., Dullemond, C.P. 2010. The outcome of protoplanetary dust growth : pebbles, boulders, or planetesimals ?. I. Mapping the zoo of laboratory collision experiments. Astronomy and Astrophysics 513. Doi:10.1051/0004-6361/200912852

[29] Gole, D.A., Simon, J.B., Li, R., Youdin, A.N., Armitage, P.J. 2020. Turbulence Regulates the Rate of Planetesimal Formation via Gravitational Collapse. The Astrophysical Journal 904. doi:10.3847/1538-4357/abc334

[30] Hansen, B.M.S. 2009. Formation of the Terrestrial Planets from a Narrow Annulus. The Astrophysical Journal 703, 1131–1140. doi:10.1088/0004-637X/703/1/1131

[31] Owen, J.E., Wu, Y. 2017. The Evaporation Valley in the Kepler Planets. The Astrophysical Journal 847. doi:10.3847/1538-4357/aa890a

[32] Izidoro, A. and 6 colleagues 2021. Formation of planetary systems by pebble accretion and migration: Hot super-Earth systems from breaking compact resonant chains. Astronomy and Astrophysics 650. doi:10.1051/0004-6361/201935336

[33] Morbidelli, A., Libourel, G., Palme, H., Jacobson, S.A., Rubie, D.C. 2020. Subsolar Al/Si and Mg/Si ratios of non-carbonaceous chondrites reveal planetesimal formation during early condensation in the protoplanetary disk. Earth and Planetary Science Letters 538. doi:10.1016/j.epsl.2020.116220

[34] Tornabene, H. A., Ash, R. D. & Walker, R. J. New insights to the genetics, formation and crystallization history of group IC iron meteorites. 52$^{nd}$ Lunar Planet. Sci. Conf., LPI Contr. No. 2548, (2021).

[35] Hilton, C.D., Ash, R.D., Walker, R.J. 2020. Crystallization histories of the group IIF iron meteorites and Eagle Station pallasites. Meteoritics and Planetary Science 55, 2570–2586. doi:10.1111/maps.13602

[36] Stewart, S.T. and 7 colleagues 2019. Collapsing Impact Vapor Plume Model for Chondrule and Chondrite Formation. Lunar and Planetary Science Conference.

[37] Lichtenberg, T., Drążkowska, J., Schönbächer, M., Golabek, G.J., Hand, T.O. 2020. Bifurcation of planetary building blocks during Solar System formation. Science 371, 6527

[38] Chiang, E., Youdin, A.N. 2010. Forming Planetesimals in Solar and Extrasolar Nebulae. Annual Review of Earth and Planetary Sciences 38, 493–522. doi:10.1146/annurev-earth-040809-152513

[39] Dullemond, C. 2013. Theoretical Models of the Structure of Protoplanetary Disks. Heidelberg University

[40] Bitsch, B., Morbidelli, A., Lega, E., Crida, A. 2014. Stellar irradiated discs and implications on migration of embedded planets. II. Accreting-discs. Astronomy and Astrophysics 564. doi:10.1051/0004-6361/201323007

[41] Kley, W., Bitsch, B., Klahr, H. 2009. Planet migration in three-dimensional radiative discs. Astronomy and Astrophysics 506, 971–987. doi:10.1051/0004-6361/200912072

[42] Kley, W. 1989. Radiation hydrodynamics of the boundary layer in accretion disks. I - Numerical methods. Astronomy and Astrophysics 208, 98–110.

[43] Rafikov, R.R. 2015. Viscosity Prescription for Gravitationally Unstable Accretion Disks. The Astrophysical Journal 804. doi:10.1088/0004-637X/804/1/62

[44] Birnstiel, T., Fang, M., Johansen, A. 2016. Dust Evolution and the Formation of Planetesimals. Space Science Reviews 205, 41–75. doi:10.1007/s11214-016-0256-1

[45] Ormel, C.W. 2014. An Atmospheric Structure Equation for Grain Growth. The Astrophysical Journal 789. doi:10.1088/2041-8205/789/1/L18

[46] Gonzalez, J.-F., Laibe, G., Maddison, S.T. 2017. Self-induced dust traps: overcoming planet formation barriers. Monthly Notices of the Royal Astronomical Society 467, 1984–1996. doi:10.1093/mnras/stx016

[47] Carballido, A., Stone, J.M., Pringle, J.E. 2005. Diffusion coefficient of a passive contaminant in a local MHD model of a turbulent accretion disc. Monthly Notices of the Royal Astronomical Society 358, 1055–1060. doi:10.1111/j.1365-2966.2005.08850.x

[48] Burkhardt, C., Dauphas, N., Hans, U., Bourdon, B., Kleine, T. 2019. Elemental and isotopic variability in solar system materials by mixing and processing of primordial disk reservoirs. Geochimica et Cosmochimica Acta 261, 145–170. doi:10.1016/j.gca.2019.07.003

[49] Brennecka, G. A. *et al.* Astronomical context of Solar System formation from molybdenum isotopes in meteorite inclusions. *Science* **370**, 837, (2020).

[50] Budde, G., Burkhardt, C. & Kleine, T. Molybdenum isotopic evidence for the late accretion of outer solar system material to Earth. *Nature Astron.* **3**, 736-741, (2019).

**FIGURES**

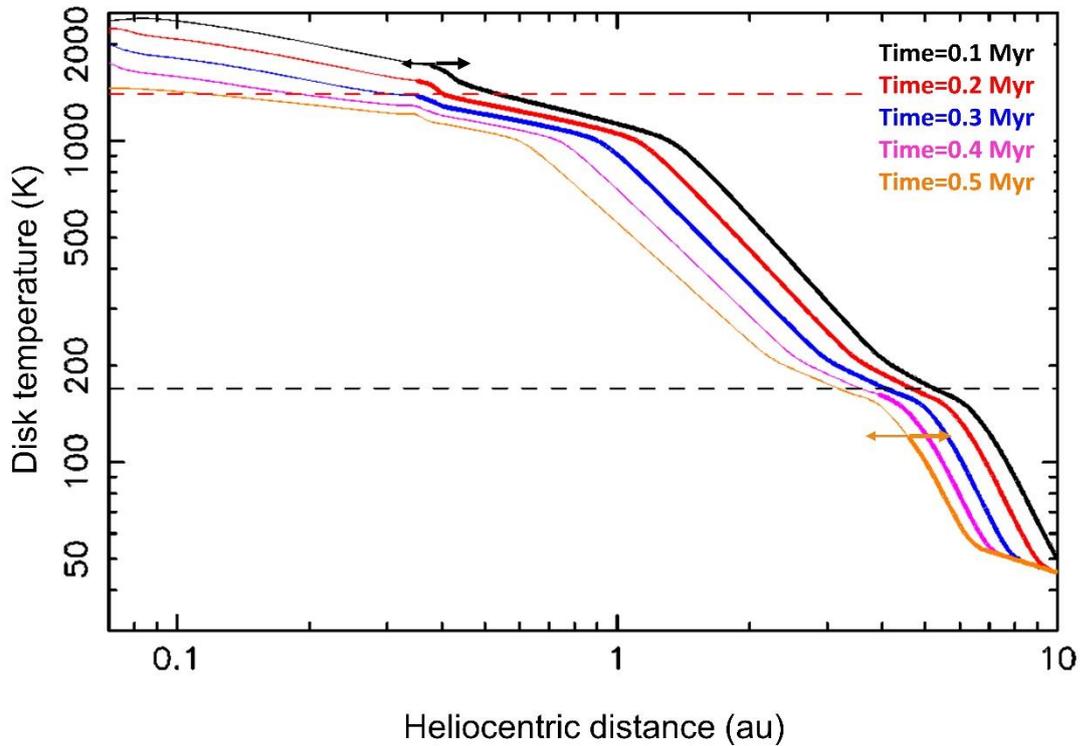

**Fig.1** The radial distribution of the disk's temperature at different times, obtained assuming $R_c=0.35\text{au}/[M_{sun}(t)]^{0.5}$. The thick part of each curve shows the region where the radial velocity of the gas is positive (outward), whereas the thin part depicts the accretion part of the disk (negative radial velocity), as also indicated by the black and orange arrows. The horizontal dashed lines mark the condensation temperature of water (T=170K, black), and rocks (T=1,400K, red). The intersection of these lines with the various colored curves identify the location of the condensation/sublimation fronts of these elements as a function of time.

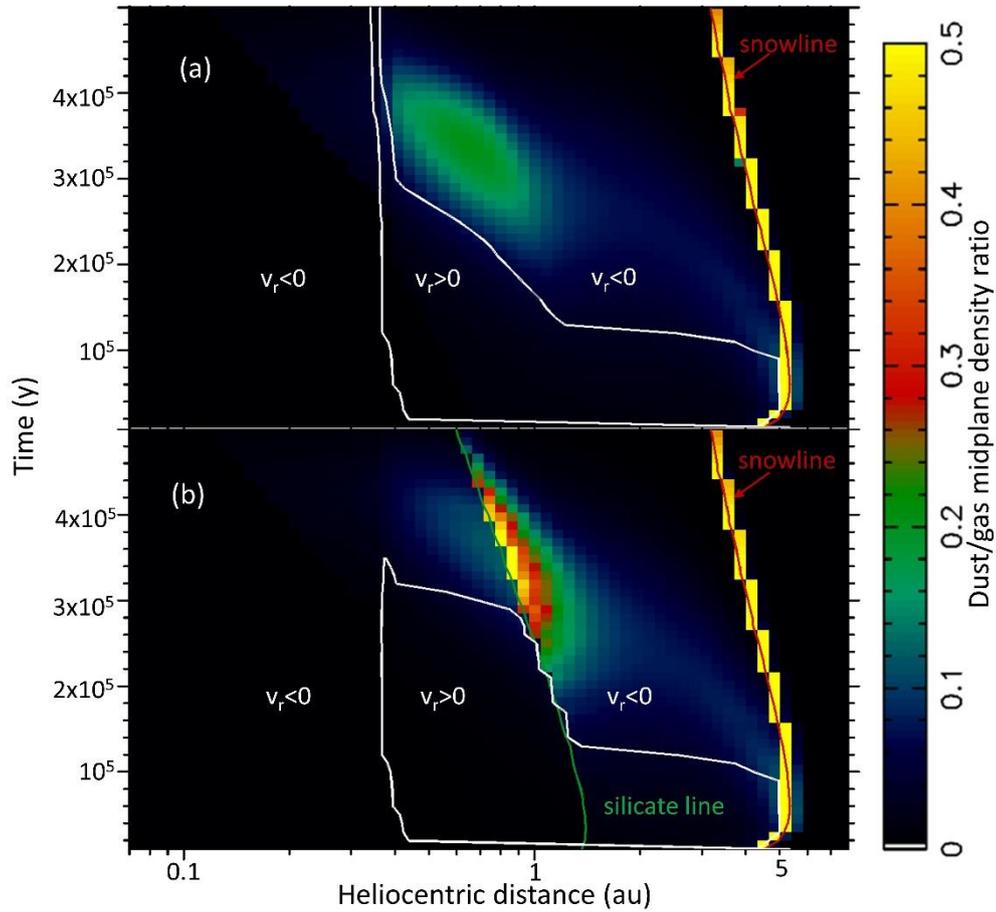

**Fig.2** The volume density ratio between dust and gas (color scale) as a function of heliocentric distance and time. Panel (a) is for the case where only the snowline is accounted for and panel (b) is for the simulation where also silicate sublimation and recondensation at T=1,000K are taken into account. The red and green curves show the location of the snowline and silicate line as a function of time. The white line separates the (r,t) domain where the radial velocity of the dust particles is negative from that where it is positive. It is different in the two panels because in (b) particles are smaller inward of the silicate line and therefore are more coupled with the gas. In (a) outward particle motion is due to the positive radial motion of the gas up to 0.29 Myr; after, there is a narrow pressure bump due to the rapid increase of viscosity with temperature that causes a drop in surface density (Extended Data Fig. 1). In (b) the particles are too small to feel the pressure bump and drift inward with the gas for t>0.35 Myr. Convergent migration and dust pile-up tend to occur at the outer white line but are contrasted by particle diffusion. When the dust/gas ratio exceeds 0.5 (yellow color) planetesimal formation as assumed to occur.

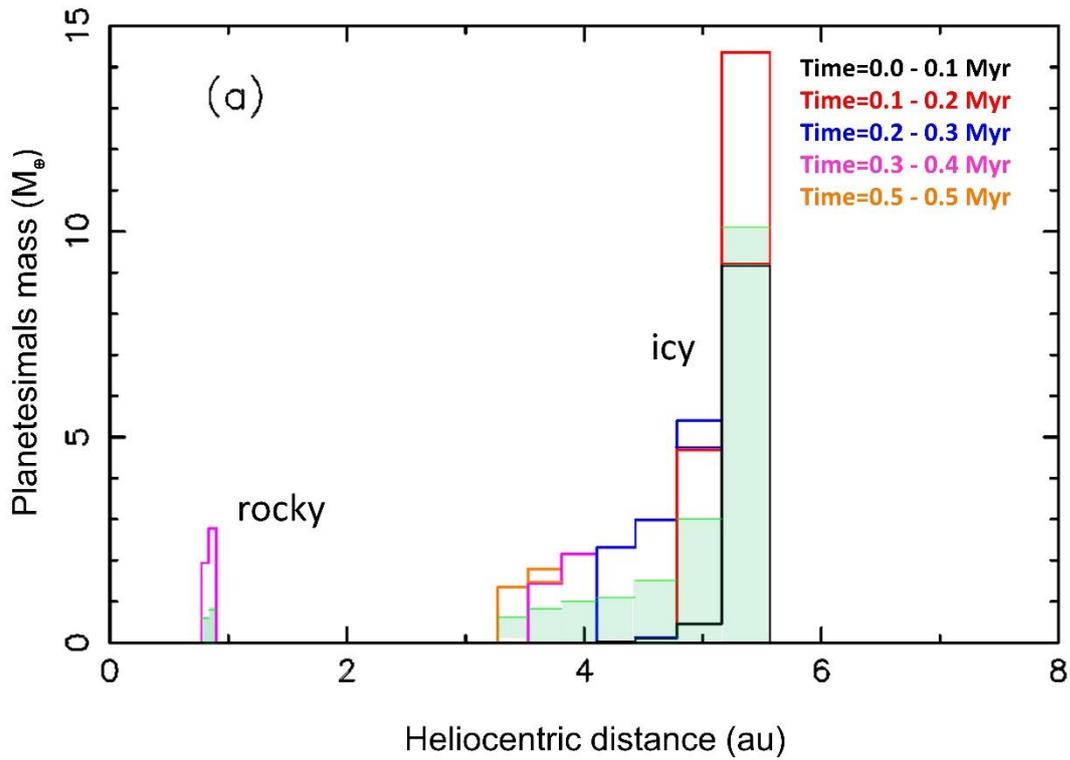

**Fig.3** The radial mass distribution of the planetesimal populations formed in different time intervals (color-coded as indicated by the labels). Planetesimal masses formed in the same radial bin at subsequent time intervals are plotted on top of each other, so that the upper border of the histogram represents the total mass ever produced. The light green shaded area shows the fraction of this mass derived from early infalling material (the material accreted onto the disk before 20 Kyr). This fraction is 46-70% for the icy planetesimals between 3.1 and 5 au and increasing from inside-out (the mass-weighted average is 60%), and 28-30% for the rocky planetesimals, just inward of 1 au.

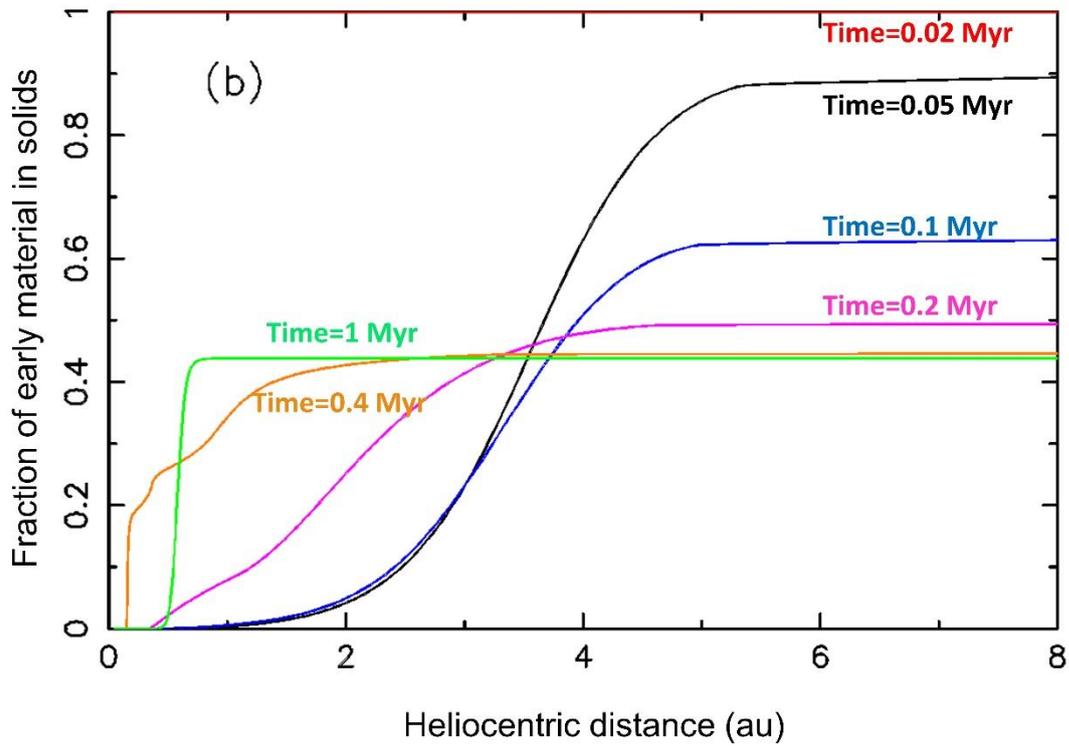

**Fig. 4** The relative proportions of early-infalling material among the disk's refractory elements, as a function of heliocentric distance at different times. By assumption, the late-infalling material starts at $t=20$ Kyr; thus up to this time 100% of the disk is made of early-infalling material. The arrival of late-infalling material chases progressively the early-infalling material from the inner disk until 0.1 Myr. Then diffusion and the inward drift of dust raises again the early/late material ratio in the inner disk, while this ratio still decreases beyond ~4 au.

# Extended Data

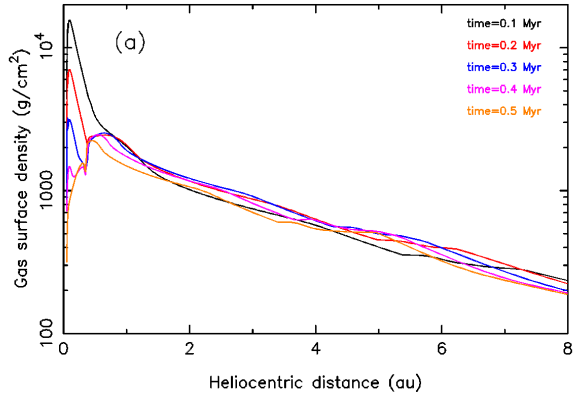

Extended Data Fig. 1. : Surface density of the disk as a function of heliocentric distance at different times.

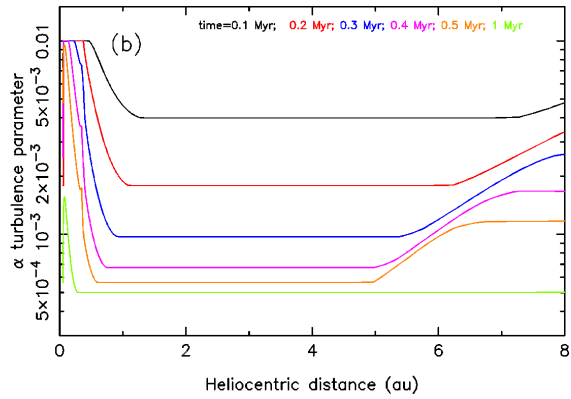

Extended Data Fig. 2. : Turbulent parameter α, for the nominal simulation presented in the main text.

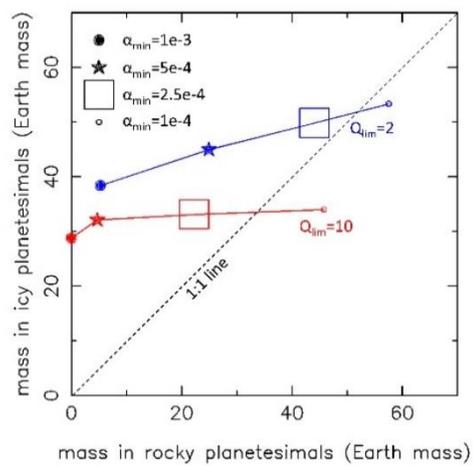

Extended Data Fig. 3: total masses of rocky and icy planetesimals for 4 values of $\alpha_{min}$ and two values of $Q_{lim}$.

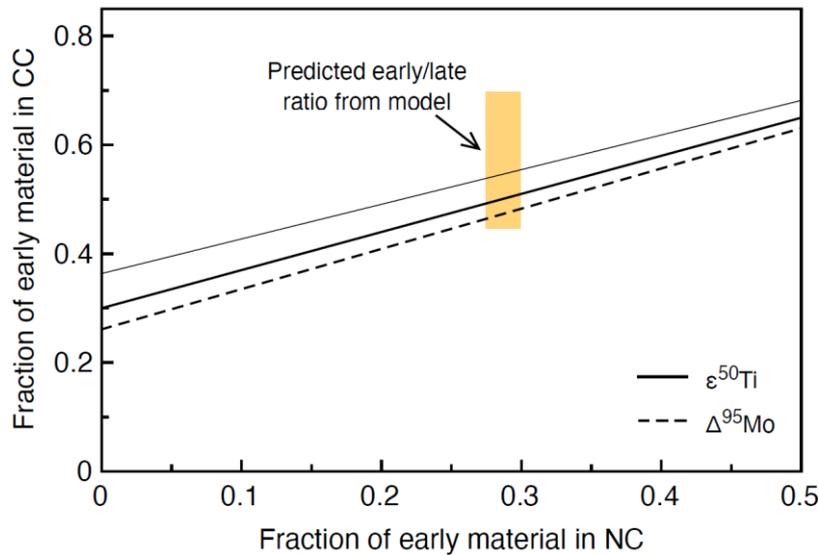

Extended Data Fig. 4: Relation between fraction of early material in CC and NC as given by Ti and Mo isotope anomalies in meteorites. The thick solid line assumes the average value for NC meteorites $\varepsilon^{50}Ti_{NC} = -1$ while the thin line assumes $\varepsilon^{50}Ti_{NC} = -2$ (i.e. the extreme value observed in NC). Orange-shaded area indicates the predicted fractions of our model: 0.275-0.3 for NC planetesimals and 0.45-0.70 for CC planetesimals.

Extended Data Table 1: HSE concentrations (ppm) for bulk iron meteorite cores and chondrites

|  | Re | Os | Ir | Ru | Pt | Reference |
|---|---|---|---|---|---|---|
| CI | 40.7 | 491 | 462 | 688 | 943 | [73] |
| Average OC | 58.4 | 679 | 585 | 880 | 1185 | [74] |
| Average EC | 55.9 | 637 | 583 | 873 | 1186 | [74] |
| *CC irons* | | | | | | |
| IIC | 280 | 3350 | 3050 | 4340 | 6070 | [75] |
| IID | 1100 |  | 10500 |  | 17400 | [76] |
| IIF | 355 | 4200 | 4200 | 6800 | 8200 | [35] |
| IVB | 2800 | 37000 | 27000 | 27400 | 29500 | [77] |
| *NC irons* | | | | | | |
| IC |  |  | 2330 |  |  | [34] |
| IIAB |  |  | 3300 |  |  | [78] |
| IIIAB |  |  | 2100 |  |  | [78] |
| IVA | 295 | 3250 | 2700 | 3900 | 5900 | [79] |

Extended Data Table 2: Core mass fractions of NC and CC iron parent bodies

| Iron meteorite group | Core mass fraction |
| --- | --- |
| *CC irons* | |
| IIC | 0.15 |
| IID | 0.05 |
| IIF | 0.11 |
| IVB | 0.02 |
| | |
| *NC irons* | |
| IC | 0.25 |
| IIAB | 0.18 |
| IIIAB | 0.28 |
| IVA | 0.21 |

**References :**

[73] Horan, M. F., Walker, R. J., Morgan, J. W., Grossman, J. N. & Rubin, A. E. Highly siderophile elements in chondrites. *Chem. Geol.* **196**, 5-20, (2003).

[74] Walker, R. J. Highly siderophile elements in the Earth, Moon and Mars: Update and implications for planetary accretion and differentiation. *Chem Erde-Geochem.* **69**, 101-125, (2009).

[75] Tornabene, H. A., Hilton, C. D., Bermingham, K. R., Ash, R. D. & Walker, R. J. Genetics, age and crystallization history of group IIC iron meteorites. *Geochim. Cosmochim. Acta* **288**, 36-50, (2020).

[76] Wasson, J. T. & Huber, H. Compositional trends among IID irons; their possible formation from the P-rich lower magma in a two-layer core. *Geochim. Cosmochim. Acta* **70**, 6153-6167, (2006).

[77] Walker, R. J. *et al.* Modeling fractional crystallization of group IVB iron meteorites. *Geochim. Cosmochim. Acta* **72**, 2198-2216, (2008).

[78] Chabot, N. L. Sulfur contents of the parental metallic cores of magmatic iron meteorites. *Geochim. Cosmochim. Acta* **68**, 3607-3618, (2004).

[79] McCoy, T. J. *et al.* Group IVA irons: New constraints on the crystallization and cooling history of an asteroidal core with a complex history. *Geochim. Cosmochim. Acta* **75**, 6821-6843, (2011).

# Supplementary information

*S1: Effects of changing key parameters*

We discuss here how the results presented for the nominal simulation in the main text change if the values of the key parameters are varied.

*S1.1*: Prescription for centrifugal radius $R_c$.
Using the usual formulation from Ref.[51] as applied in previous studies[17-19], i.e. $R_c(t)$=10au $(M_{sun}(t)/M_\odot)^3$, the centrifugal radius expands very rapidly and the gas radial velocity becomes negative in the inner part of the disk in just 10 Kyr. Consequently, there is no pile-up of dust near 1au, even if $\alpha_{min}$ is reduced to $10^{-4}$ (as discussed below, a smaller value of $\alpha_{min}$ enhances planetesimal production). The sublimation/recondensation effect alone is not enough to reach the required critical dust/gas ratio for planetesimal formation. Consequently, planetesimals form only at the snowline. Assuming $R_c(t)=R_{c(0)}/(M_{sun}(t))^{0.5}$ as in the nominal simulation, but with $R_{c(0)}$=1au also does not allow the formation of planetesimals at the silicate-sublimation line (also for $\alpha_{min}=10^{-4}$) because in this case the radial velocity of the gas is never positive inward of 1.06au. Thus, it is essential that the centrifugal radius is significantly smaller than the distance where silicates sublimate in order to have a positive gas radial velocity and allow planetesimal formation at that location. The magnitude of the positive velocity also has importance. For instance, assuming $R_c(t)$=0.35au/$M_{sun}(t)$ (constant angular momentum of the infalling material), rocky planetesimal formation is significantly reduced because initially the gas falls at a larger distance (at t=0, when $M_{sun}(0)$= ½ $M_\odot$, $R_c(0)$ is bigger), so that the rate of radial expansion of the gas at the silicate sublimation line is reduced and turns negative ~20 Kyr earlier. The efficiency of rocky planetesimal formation can be restored to values similar to those of the nominal simulation by reducing the viscosity $\alpha_{min}$ in the disk.

It is also important that the positive radial velocity of the gas lasts long enough. For instance, if the infall rate is adjusted so that the star-disk system reaches $1M_\odot$ in 170 Kyr and then is truncated as in Ref.[17-19], even assuming $R_c(0)$=0.35 au there would be no planetesimal formation near 1 au because the radial velocity of the gas turns negative as soon as the infall is suppressed, too early for a significant dust pile-up to be generated.

A small centrifugal radius requires efficient magnetic breaking to remove most of the angular momentum of the infalling gas. The efficiency of magnetic breaking depends on the intensity of the magnetic field, its coupling with the gas, the strength of ambipolar diffusion. Different simulations adopting different parameters depict different results. Ref.[52] shows streamers of gas feeding the disk at large distances from the central star, while in Ref.[27] the material falls very close to the central sink. It is possible that the situation may be different in reality from case to case. If magnetic breaking is not very efficient, very extended and massive disks form, which are prone to develop gravitational instabilities[52], while in presence of efficient breaking the resulting disks are rather small[27,53] that are never gravitationally unstable. Both cases are observed in the collection of extrasolar disks. Considerations based on the orbital distribution of trans-Neptunian objects suggest that the circumsolar disk was rather small, with a radial extension smaller than ~80 au[54]. Thus, it is reasonable to expect that magnetic breaking was strong, in the protosolar case.

*S1.2*: Prescription for $\alpha$.
Two parameters are important in our prescription of the viscosity parameter: $\alpha_{min}$ and $Q_{lim}$. We have tested $\alpha_{min}=10^{-4}$ and, with the same Schmidt number Sc=10, we obtained an enhanced production of planetesimals near 1au, exceeding 30 $M_\oplus$. One could reduce Sc (proportionally to the ratio of $\alpha$ values at the time of rocky planetesimal formation in the nominal simulation, i.e. ~0.4 Myr) to reduce the total rocky planetesimal mass. Nevertheless, the two simulations (that with $\alpha_{min}=10^{-4}$, Sc=4 and the nominal one with $\alpha_{min}=5\times10^{-4}$, Sc=10) would not be equivalent: the disk would be colder than in the nominal simulation because of the reduced viscous heating; moreover the product $\alpha$Sc at $t<0.4$ Myr would be larger. The first effect shifts the ring of rocky planetesimals sunwards and the second effect reduces the amount of icy planetesimals formed at 5au, in favor of those formed at 3-4 au. The new distribution of planetesimals is shown in Fig.S1. Notice also that planetesimal formation at 1 au occurs ~0.1 Myr later than in the nominal simulation.

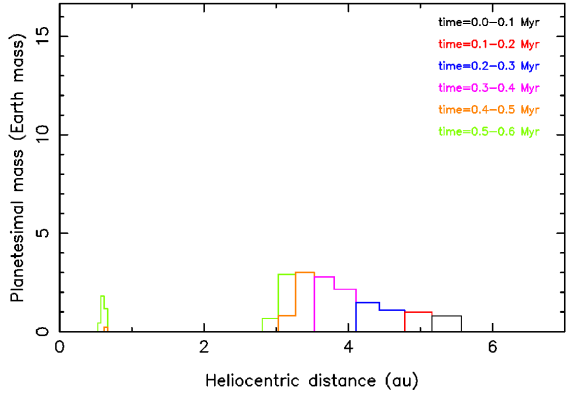

Fig S1: The same as Fig.3 of the main text, but changing $\alpha_{min}$ to $10^{-4}$ and Sc to 6. Moreover, no information on the fraction of "early" disk material is provided here.

Concerning the value of $Q_{lim}$ we tested the case $Q_{lim}=2$ recommended in Ref.[43]. A portion of the central disk (that with $2<Q<10$) becomes less viscous more rapidly and consequently the gas has a reduced propensity to spread radially. The surface density of the disk is higher in the outer part and smaller in the inner part of the disk, and remains flat in the 3-7 au region (Fig. S2a). At $t=1$ Myr there is still a density of gas of $\sim500$g/cm$^2$ at 6 au, which is about twice the value we find in our nominal simulation. The simulation with $Q_{lim}=2$ produces a much larger mass of planetesimals both at the silicate-sublimation line and at the snowline (Fig. S2b). The former is due to the fact that there is less gas in the inner part of the disk, so that the particles' Stokes number is bigger; the latter is due to the reduced viscosity in the snowline region.

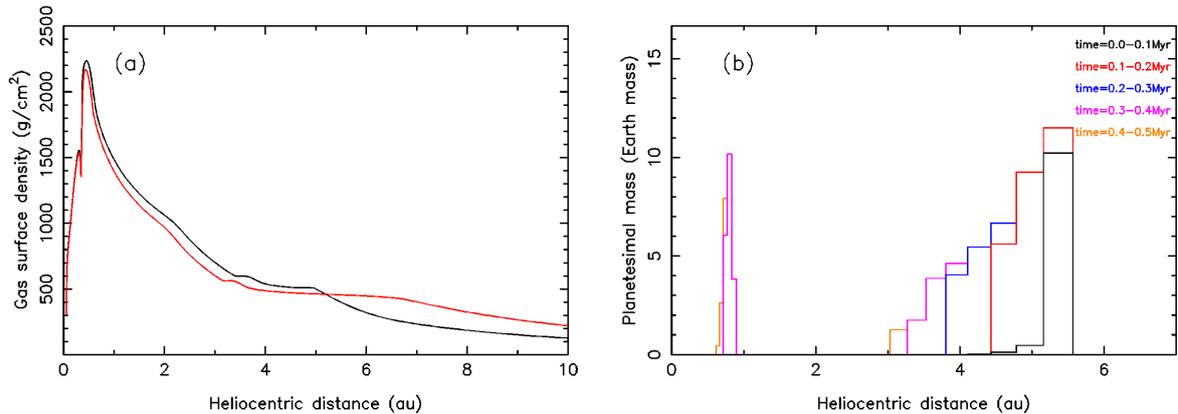

Fig. S2: (a) comparison between the disk's surface density distributions at $t=0.5$ Myr between the nominal simulation (black) and that with $Q_{lim}=2$ (red). (b) the planetesimals' radial mass distribution in the simulation with $Q_{lim}=2$. Compare with Fig.3 of the main text. No information on the fraction of "early" disk material is provided here.

*S1.3*: Planetesimal formation rate.
We compared the prescriptions given in Ref.[11] (nominal simulation) and Ref.[13] for the rate at which dust is converted into planetesimals when once $\rho_d/\rho_g>0.5$. Assuming $\varepsilon=0.1$ in the prescription of Ref.[13] suppresses the formation of rocky planetesimals, while it doubles the total mass of icy planetesimals. However, it is enough to reduce $\alpha_{min}$ to recover rocky planetesimal formation. For $\alpha_{min}=10^{-4}$ we obtain a total mass in rocky planetesimals that is 3 times that of our nominal simulation.

*S1.4*: Dependence on the planetesimals' total masses on parameters.
Extended Data Fig. 3 shows the total masses of icy and rocky planetesimals as a function of $\alpha_{min}$ for $Q_{lim}=2$ and 10. The total mass in rocky planetesimals increases sharply with decreasing $\alpha_{min}$ while the total mas sin icy planetesimals is less affected. This is because in our model most of icy planetesimals form early, when $\alpha$ is still much larger than $\alpha_{min}$, particularly in the case $Q_{lim}=10$ in which the value of $\alpha$ is set by $Q$ and not by the amount of material infalling onto the disk. Notice from Extended Data Fig. 3 that, while it is possible to have icy planetesimals and no rocky planetesimals (large $\alpha_{min}$), the opposite is not true. Supposing that a population of planetesimals always forms a planet with a proportional mass, our model suggests that a rocky terrestrial or super-Earth planet should always be accompanied by an icy super-Earth or a giant planet (if the icy super-Earth becomes a seed for gas accretion). Observations seem to show a positive correlation between close-in super-Earths and distant giants[55,56]. The correlation between rocky super-Earths and more distant icy super-Earths is not confirmed because of the difficulty to detect distant super-Earths, but is a prediction of our model.

*S1.5*: Isotopic dichotomy.
The value of the time $t_{dich}$ at which we switch from tracer #1 to tracer #2 affects the results in a straightforward manner. If we increase $t_{dich}$ tracer #1 comes into the disk for a longer time. Hence its abundance relative to tracer #2 increases in both the icy and rocky planetesimals. The opposite is true if we decrease $t_{dich}$. A dichotomy of compositions between the two planetesimal populations would still be present but, if we fix the isotopic composition of tracer #1 to that carried by CAIs, the final isotopic properties of both CC and NC planetesimals would be less consistent with the measurements. Specifically, increasing $t_{dich}$ shifts the yellow rectangle in Extended Data Fig. 4 to the right and above the solid and dash lines.

*S1.6*: Dust maximal sizes.
In our model we assume different maximal sizes for particles on opposite sides of the snowline and silicate-sublimation line. If we eliminate the size contrast at the snowline by setting the size of the rocky particles to 10cm (which is inconsistent with the current understanding of the fragmentation and bouncing barriers of silicate dust[8]) there is still planetesimal formation beyond the snowline, but with a reduced total mass (about 20 $M_\oplus$). If instead we eliminate the size contrast by reducing the size of icy particles to 5mm, planetesimal formation at the snowline is completely suppressed (also for $\alpha_{min}=10^{-4}$). This is not just because of the lack of size contrast on the two sides of the snowline, but also (and mostly) because of the very reduced Stokes number of the icy particles. A similar experiment, suppressing the size contrast at the silicate-sublimation line by reducing the size of silicate particles to 1mm, resulted in no planetesimal formation near 1au (also for $\alpha_{min}=10^{-4}$). Instead, if this size contrast is suppressed by increasing the size of refractory particles to 5mm, the formation of rocky planetesimals is impeded for $\alpha_{min}=5\times10^{-4}$ but is recovered for smaller $\alpha_{min}$. Thus, the size contrast between particles on opposite sides of a condensation line plays an important role and appears quite crucial for the formation of rocky planetesimals.

We also assumed more sophisticated recipes for the maximal size of particles. In one, we assumed that dust can grow only until its velocity dispersion $dv=(3\alpha/Sc\ St)^{1/2}$ reaches a threshold value. We nevertheless limit the maximal size of particles to be 10cm, 5mm and 1mm in the icy, silicate and refractory regimes, if the size limit provided by the velocity dispersion threshold is less stringent. We tested velocity thresholds of 1, 3 and 5 m/s. The limits at 3 and 5 m/s lead to qualitatively similar results. The total mass in icy planetesimals is reduced, by a factor up to 2, because the solid particles at the snowline are smaller than 10cm; instead the total mass of rocky planetesimals is increased by up to 75% because (i) the size limits of 5 and 1 mm are more severe than those given by the dispersion velocity and (ii) more material avoided to be trapped into icy planetesimals and drifted into the inner disk. However, assuming a threshold velocity of 1 m/s changes the results qualitatively. Planetesimal formation at the snowline is delayed because solid particles can become big enough only at a late time, when α has decreased sufficiently. Thus icy planetesimals form only when the snowline is at 3-4 au and their total mass is reduced to ~1/3. Rocky planetesimal formation still occurs, but their total mass is reduced to ~1/10 (it could be increases by lowering $\alpha_{min}$). In the other recipe, we attempted to account for the new result that the sticking properties of ice depend on temperature[57]. Thus, we assumed that a maximal size $D_{max}=10cm\ (T/170K)^n$ for icy particles, with n=1,2,4. The result in terms of planetesimal formation does not change significantly. The particle size changes beyond the snowline, but because the temperature decays as $r^{-1/2}$ in the irradiation dominated regime, even for n=2 the decay in particle size is not very strong.

*S1.7*: Sublimation/recondensation prescription at the snowline.

In our work we assume complete evaporation of water above a fixed sublimation temperature (170K here). In reality water-vapor can co-exist with water-ice up to a maximum pressure, that depends on local temperature, the "saturating vapor pressure" [8,10,12]. So water vapor can subsist also beyond the snowline and some icy grain can survive also inward of the snowline. In principle this may change the surface density distribution of solids in the vicinity of the snowline, affecting planetesimal formation. To test the differences in the results between the two approaches, we use the code developed in Ref.[12] which accounts for partial pressure of vapor in the condensation/sublimation process, an option which can easily be removed for comparisons. The disk modeled in Ref.[12] is different from the one modeled here: it is colder and less massive, so that the snowline is much closer to the central star. Nevertheless we can simulate a situation analogue to the one investigated here if the diffusion coefficient in normalized coordinates at the snowline is the same in the two cases. The diffusion coefficient in normalized coordinates (r=1, Ω=1) is $D=\alpha h^2/Sc$, where h is the disk's aspect ratio and Sc is the Schmidt number. For a given temperature (i.e. T=170K at the snowline) h is proportional to $r_{snow}^{1/2}$. In the code of Ref. [12] we assume $\alpha=10^{-3}$ and Sc=1; the snowline is at $r_{snow}$=1.4 au (h=0.037), and therefore D=1.37x$10^{-6}$. We find the same in our disk at t=150,000y, when $r_{snow}$=5 au (h=0.07), $\alpha$=2.8x$10^{-3}$ and Sc=10. Notice that at this time icy planetesimal formation is fully under way in our model.

Fig. S3 shows the results provided by the code in Ref.[12] if instantaneous sublimation/condensation is assumed (left panel) or the accurate calculation based on water vapor partial pressure is implemented (right panel). In the left panel, there is an abrupt transition between the water vapor density (dotted blue line) and icy grains density (solid blue line). In the right panel, the density of water vapor extends beyond the snowline (dashed-blue line). Nevertheless the density distribution of icy solids is very similar in both cases. This is because there is nevertheless a one-order-of-magnitude drop in the density of vapor at the snowline and the grains condensed there are allowed to diffuse radially.

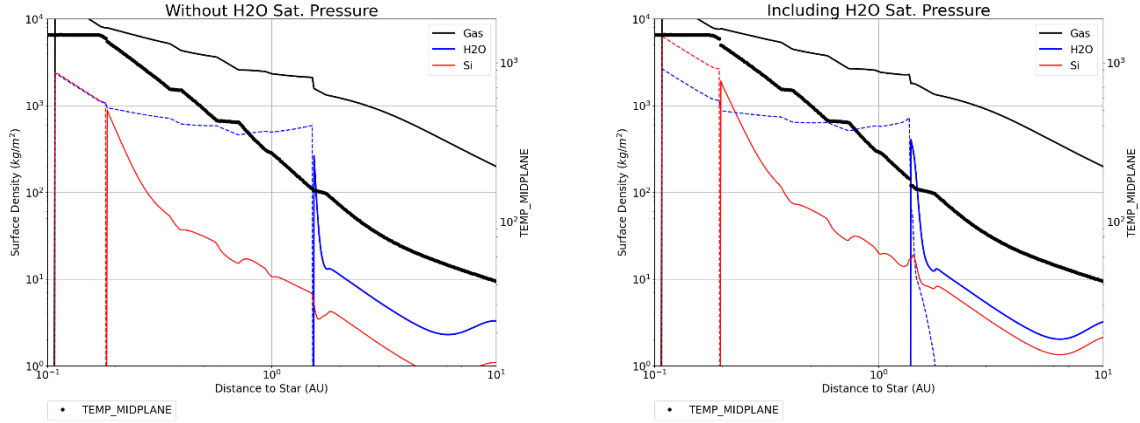

Fig. S3. The surface density of the disk (black thick line), of water (blue line: dotted for vapor and solid for ice) and silicate (red line) and temperature of the disk (black thin line) in the model of Ref.[12]. In this experiment the diffusion coefficient at the snowline is the same as in our model during icy planetesimal formation. In the left panel water vapor is turned instantaneously to solid at the snowline, while in the right panel water sublimation/recondensation is computed taking into account the partial pressure of water vapor. The distribution of ice turns out to be almost identical in the two cases, indicating that in case of significant diffusion the instantaneous recondensation of water at the snowline is a good approximation.

The situation would be different if D were much smaller (e.g. for $\alpha=10^{-4}$); in this case the distribution of solids would be less peaked beyond the snowline if the condensation rate were computed from the water vapor partial pressure. However, planetesimal formation is over in our model before that these conditions are met. Thus, we conclude that our simplified treatment of evaporation/condensation is adequate for our purposes.

*S2: A trade-off between viscosity and gas density*

The formation of planetesimals requires to find a sweet-spot in the $(\alpha, \sum_g)$ parameter space. A small value of $\alpha$ helps the sedimentation and the radial concentration of particles, but does not spread the gas of the disk efficiently. Thus $\sum_g$ decays over time more slowly and the larger density of gas reduces the particles' Stokes number, vanishing the positive effects of the reduced viscosity. A larger value of $\alpha$ reduces the density of gas, which enhances the particles' Stokes number, but increases the particle vertical stirring and radial diffusion. For these reasons it is important that $\alpha$ is large at the beginning of the disk's evolution -so to favor rapid disk spreading and density decay- and then decreases to small values, as in our model. A simulation where $\alpha=10^{-2}$ throughout the simulation as in Ref.[17-19] would not produce planetesimals, unless an extreme value of Sc is adopted, as in Ref.[9]. However, remember that $\alpha$ cannot become too small, otherwise the disk becomes too cold to form planetesimals near 1 au (given that the formation site of rocky objects is related to the location of the silicate sublimation line), unless a heating mechanism operates in the disk in addition to viscous dissipation. Ohmic dissipation may be such a mechanism[58].

*S3: Condensation temperatures and the formation of refractory-rich bodies*

In this work we have invoked the sublimation/recondensation of 50% of the non-volatile material (dubbed generically as *silicates*) at *T*=1,000K. In a previous publication[33] we found evidence in the meteoritical record for a similar separation of materials, that we called

*refractory materials* and *residual condensates,* but with condensation temperatures above and below 1,400K respectively. The recondensation of silicate vapor beyond the *T*=1,000K line in this work reproduces the process of formation of the residual condensates invoked in Ref.[33] (see also Ref.[22]), but the temperatures (1,000K vs. 1,400K) do not match. The mismatch is more apparent than real because Ref.[33] showed that the temperature at which refractory elements and residual condensates need to be separated depends on the pressure in the disk and its C/O ratio; they showed it could be 1,060K for $P=10^{-4}$ bar and C/O=1.0. In this work, if we had assumed a larger temperature for the sublimation of silicates, the ring of rocky planetesimals would have shifted somewhat towards the Sun. The new location can be deduced from the intersection of the disk's temperature curve in Fig. 1 for *t*=0.4 Myr with the required value of the temperature. Thus, restoring the ring near 1 au would require a hotter disk than the one produced in the reference simulation, possibly due to Ohmic dissipation[68].

A more conceptual difference is that Ref.[33] argued that the refractory grains formed refractory-rich planetesimals which ultimately contributed to forming a refractory-rich Earth. In this work, instead, the refractory grains never reach a solid/gas ratio large enough to trigger planetesimal formation. This is because the pressure bump generated by the drop in gas surface density in the inner part of the disk (Extended Data Fig. 1, S2a), associated with the increase in viscosity (Extended Data Fig. 2), is not sharp enough to trap mm-sized particles[59]. If this is correct, explaining the formation of a refractory-rich Earth requires invoking the evaporation of Si from warm planetary embryos with an initial enstatite chondrite-like (i.e. supra-solar) Si/refractory-element ratio[60].

*S4: Formation of CAIs and other refractory condensates*

In our model, following Ref.[23], we identify CAIs with condensates from early-infalling material. This identification is necessary to explain why CAIs have a solar isotopic composition for oxygen. Following this identification, our model implies that CAI formation lasted up to $t_{dich}$=20 Kyr. It has been argued that the CAI formation period may have last up to[71] 200 Kyr, but this prolonged interval likely includes later reprocessing of CAIs following their formation. Consistent with this, a bulk Al-Mg isochron for CAIs, as well as internal Al-Mg isochrons for the most primitive CAIs are consistent with a much shorter formation interval of ~20 Kyr or less[61].

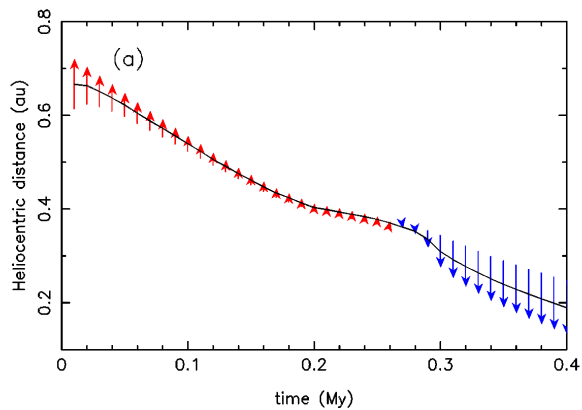

Fig.S4: The black curve shows the radial location where T=1,400K as a function of time. The arrows depict the radial displacement of the gas over 400y intervals. When arrows are red the gas cools during its radial motion, so condensation of refractory minerals is possible. The opposite is true when arrows are blue. The timescale over which condensation occurs (here 0.2 Myr) increases (decreases) if the assumed value of the centrifugal radius $R_c(0)$ is decreased (increased).

However, in our model the condensation of refractory minerals continues as long as the gas has a positive radial flow across the $T$=1,400 K line, i.e. for 0.2 Myr (Fig.S4). Thus our model implies the existence of grains with a CAI-like chemical composition but a NC isotopic composition. Refractory grains are not identified directly in NC chondrites, but they are expected to have been reprocessed in the formation of Al-rich chondrules. The isotopic analysis of these chondrules[62] revealed no isotopic anomalies pointing towards those of CAIs. This suggests that refractory grains with a NC isotopic composition indeed formed in the inner disk, in agreement with our model.

*S5: Formation of late planetesimals: an outlook*

This paper addresses the formation of early planetesimals, accreted in the first ½ Myr and related to iron meteorite parent bodies. The meteorite record, however, shows that there are also planetesimals that formed significantly later, such as the parent bodies of chondrites, whose formation times are constrained to be at least 2 Myr after CAI by the analysis of the ages of individual chondrules[63]. Like the parent bodies of all meteorites, those of chondrites are today in the asteroid belt, but originally they may have formed elsewhere, e.g. near 1 au for the NC chondrites[64] and beyond Jupiter for the CC chondrites[65,66,16], i.e. more or less in the same place where our early planetesimals form. The Kuiper belt objects also appear to have formed late, given that those with diameter D<700 km have a low bulk density implying the lack of internal differentiation[67]; they definitely formed beyond the location of our first icy planetesimals, up to about 45 au. Our model is not appropriate to discuss the formation of late planetesimals because we assume that all the dust in a radial bin has a unique size. Due to the lack of small dust strongly coupled with the gas, all the solid material in our model drifts towards the Sun quite rapidly. By 1 Myr, basically all the particles that have not been incorporated into planetesimals have drifted to the inner edge of the disk. Thus, there is no material left to form late planetesimals. Studying the formation of late planetesimals requires to consider that a significant fraction of the mass remains for long time stored in small particles that have a limited radially drift, as in Ref.[68]. It also requires to account for the formation of Jupiter, to block the radial drift of outer solar system particles as they grow in size and the photo-evaporation of the gas[69], in order to eventually increase the dust/gas ratio above the planetesimal-formation threshold. All these features are not present in our model and will be the object of future developments.

The formation of NC chondrites is even more complex. The fact that NC chondrites and NC irons have very similar isotopic properties requires little-to-no contamination from CC dust over millions of years. One possibility is that Jupiter forms at the same time as the NC iron meteorite parent bodies (~0.4 Myr in our model). In this case, the early/late material ratio in the inner solar system would be frozen at the NC value because the Jupiter's barrier prevents the penetration of new CC dust from the outer disk. However, the preservation of the material not incorporated in the first NC planetesimals until the chondrite formation time is problematic. In a low viscosity disk, Jupiter may form rings inwards of its orbital radius[70], possibly helping the preservation of dust. Another possibility is that the material that forms the NC chondrites is generated as debris in collisions among the first NC planetesimals. There is indeed a growing literature on the possibility that chondrules are collisional debris[36,71,72]. In this case the isotopic similarity between NC chondrites and irons would be obvious, because the two are genetically linked. In this case Jupiter would not need to form at the same time of NC iron meteorite parent bodies. It would just need to form at a generic time prior to NC chondrite formation, in order to keep the inner solar system clean of CC dust when such formation happened.

**Supplementary references**

[51] Shu, F.H. 1977. Self-similar collapse of isothermal spheres and star formation. The Astrophysical Journal 214, 488–497. Doi :10.1086/155274

[52] Kuffmeier, M., Frimann, S., Jensen, S.~S., Haugbolle, T. 2018. Episodic accretion: the interplay of infall and disc instabilities. Monthly Notices of the Royal Astronomical Society 475, 2642–2658. doi:10.1093/mnras/sty024

[53] Hennebelle, P., Commercon, B., Lee, Y.-N., Charnoz, S. 2020. What determines the formation and characteristics of protoplanetary discs?. Astronomy and Astrophysics 635. doi:10.1051/0004-6361/201936714

[54] Kretke, K.A., Levison, H.F., Buie, M.W., Morbidelli, A. 2012. A Method to Constrain the Size of the Protosolar Nebula. The Astronomical Journal 143. doi:10.1088/0004-6256/143/4/91

[55] Zhu, W., Wu, Y. 2018. The Super Earth-Cold Jupiter Relations. The Astronomical Journal 156. doi:10.3847/1538-3881/aad22a

[56] Bryan, M.L. and 6 colleagues 2019. An Excess of Jupiter Analogs in Super-Earth Systems. The Astronomical Journal 157. doi:10.3847/1538-3881/aaf57f

[57] Musiolik, G., Wurm, G. 2019. Contacts of Water Ice in Protoplanetary Disks-Laboratory Experiments. The Astrophysical Journal 873. doi:10.3847/1538-4357/ab0428

[58] Béthune, W., Latter, H. 2020. Electric heating and angular momentum transport in laminar models of protoplanetary discs. Monthly Notices of the Royal Astronomical Society 494, 6103–6119. doi:10.1093/mnras/staa908

[59] Ueda, T., Flock, M., Okuzumi, S. 2019. Dust Pileup at the Dead-zone Inner Edge and Implications for the Disk Shadow. The Astrophysical Journal 871. doi:10.3847/1538-4357/aaf3a1

[60] Young, E.D. and 6 colleagues 2019. Near-equilibrium isotope fractionation during planetesimal evaporation. Icarus 323, 1–15. doi:10.1016/j.icarus.2019.01.012

[61] MacPherson, G.J., Kita, N.T., Ushikubo, T., Bullock, E.S., Davis, A.M. 2012. Well-resolved variations in the formation ages for Ca-Al-rich inclusions in the early Solar System. Earth and Planetary Science Letters 331, 43–54. Doi :10.1016/j.epsl.2012.03.010

[62] Ebert, S. and 6 colleagues 2018. Ti isotopic evidence for a non-CAI refractory component in the inner Solar System. Earth and Planetary Science Letters 498, 257–265. doi:10.1016/j.epsl.2018.06.04

[63] Villeneuve, J., Chaussidon, M., Libourel, G. 2009. Homogeneous Distribution of 26Al in the Solar System from the Mg Isotopic Composition of Chondrules. Science 325, 985. doi:10.1126/science.1173907

[64] Raymond, S.N., Izidoro, A. 2017. The empty primordial asteroid belt. Science Advances 3, e1701138. doi:10.1126/sciadv.1701138

[65] Walsh, K.J., Morbidelli, A., Raymond, S.N., O'Brien, D.P., Mandell, A.M. 2011. A low mass for Mars from Jupiter's early gas-driven migration. Nature 475, 206–209. doi:10.1038/nature10201

[66] Raymond, S.N., Izidoro, A. 2017. Origin of water in the inner Solar System: Planetesimals scattered inward during Jupiter and Saturn's rapid gas accretion. Icarus 297, 134–148. doi:10.1016/j.icarus.2017.06.030

[67] Brown, M.E. 2013. The Density of Mid-sized Kuiper Belt Object 2002 UX25 and the Formation of the Dwarf Planets. The Astrophysical Journal 778. doi:10.1088/2041-8205/778/2/L34

[68] Charnoz, S., Taillifet, E. 2012. A Method for Coupling Dynamical and Collisional Evolution of Dust in Circumstellar Disks: The Effect of a Dead Zone. The Astrophysical Journal 753. doi:10.1088/0004-637X/753/2/119

[69] Carrera, D., Gorti, U., Johansen, A., Davies, M.B. 2017. Planetesimal Formation by the Streaming Instability in a Photoevaporating Disk. The Astrophysical Journal 839. doi:10.3847/1538-4357/aa6932

[70] Bae, J., Nelson, R.P., Hartmann, L. 2016. The Spiral Wave Instability Induced by a Giant Planet. I. Particle Stirring in the Inner Regions of Protoplanetary Disks. The Astrophysical Journal 833. doi:10.3847/1538-4357/833/2/126

[71] Johnson, B.C., Minton, D.A., Melosh, H.J., Zuber, M.T. 2015. Impact jetting as the origin of chondrules. Nature 517, 339–341. doi:10.1038/nature14105

[72] Choksi, N., Chiang, E., Connolly, H.C., Gainsforth, Z., Westphal, A.J. 2021. Chondrules from high-velocity collisions: thermal histories and the agglomeration problem. Monthly Notices of the Royal Astronomical Society 503, 3297–3308. doi:10.1093/mnras/stab503